\documentclass[10pt,twocolumn]{revtex4-2}
\usepackage{amssymb}
\usepackage{graphicx}
\usepackage{subcaption}
\usepackage{float}
\usepackage{bm}
\pdfoutput=1
\usepackage{epstopdf}
\usepackage[T1]{fontenc}
\usepackage{selinput}
\SelectInputMappings{%
  aacute={‡},
  ntilde={–},
  Euro={Û}
}
\usepackage{babel}

\begin{document}

\title{Materia cu‡ntica en cavidades de alta reflectancia (Many-body CQED)}
\author{Santiago F. Caballero-Ben\'{\i}tez\footnote{Corresponding author: scaballero@fisica.unam.mx}  
 }

\affiliation{ 
Instituto de F\'{\i}sica, LSCSC-LANMAC, Universidad
Nacional Aut\'onoma de M\'exico, Apdo. Postal 20-364, M\'exico D.
F. 01000, M\'exico. }

\date{\today}
\begin{abstract}
En este art'culo se discuten algunos detalles del curso sobre ``Materia cu‡ntica en redes —pticas y cavidades de alta reflectancia (Many-body CQED)'' 
de la escuela de verano de F'sica XVIII (2021) en la UNAM. Se describen conceptos œtiles para la simulaci—n cu‡ntica an‡loga, as' como se introducen detalles de modelos de materia cu‡ntica en redes —pticas y en cavidades de alta reflectancia. 
\end{abstract}

\maketitle
Las notas del curso son una introducci—n sobre la motivaci—n de la simulaci—n cu‡ntica an‡loga y su relaci—n con modelos t'picos de estudio con la siguiente estructura:
\begin{itemize}
\item{Redes —pticas y materia condensada}
\item{El modelo de Bose-Hubbard}
\item{Modelos en cavidades de alta reflectancia}
\item{Conclusiones}
\end{itemize}

\section{Redes —pticas y materia condensada}
A partir de los avances en el control de sistemas at—micos de manera experimental desde finales de la la dŽcada de los 90, se ha logrado poder enfriar a temperaturas muy cercanas al cero absoluto (del orden de $10^{-6}-10^{-9}$K) ‡tomos en sistemas al alto vac'o. En estas condiciones es posible atrapar a los ‡tomos de muy baja energ'a en estructuras peri—dicas generadas por luz de l‡seres\cite{Lewenstein}. A estas estructuras peri—dicas se les denomina redes —pticas (``optical lattices'') \cite{Jaksch}. Los ‡tomos en estas redes —pticas se comportan de manera muy similar a c—mo se comportan los electrones en un s—lido. Esto da origen a la idea de utilizar sistemas at—micos como simuladores potenciales de materiales, donde a diferencia de un material real, todos los par‡metros que dan forma a los modelos efectivos pueden ser controlados con alta precisi—n de manera externa experimentalmente\cite{Bloch, Schafer}. Desde el punto de vista te—rico, esto abre la posibilidad a la exploraci—n de nuevos modelos, que son experimentalmente realizables donde se pueden explorar mecanismos en bœsqueda de nueva fenomenolog'a y en varias instancias encontrar nuevas maneras en la materia se organiza en el rŽgimen cu‡ntico. Esto da lugar a nuevos estados de la materia, los cuales tienen el potencial de ser utilizados para el desarrollo de tecnolog'a, ya sea en su realizaci—n en el sistema at—mico o en un sistema an‡logo del estado s—lido. Estas y otras aplicaciones de la materia cu‡ntica forman parte de las ``Tecnolog'as Cu‡nticas''. En ellas se busca utilizar efectos y propiedades de sistemas en el rŽgimen cu‡ntico para el desarrollo tecnol—gico\cite{QSim}.  

Concretamente,  la analog'a entre los ‡tomos en redes —pticas y el sistema electr—nico se puede entender partir de la representaci—n de los modelos de enlace fuerte electrones. Un modelo simple t'pico efectivo de electrones en un s—lido es el modelo de Hubbard\cite{Coleman}:

\begin{equation}
\mathcal{H}^e=-t\sum_{\sigma\in\{\uparrow,\downarrow\}}\sum_{\langle i,j\rangle}(\hat c^\dagger_{i,{\sigma}}\hat c^{\phantom{\dagger}}_{j,{\sigma}}+\mathrm{H.c.})+U\sum_i\hat{n}_{i,{\uparrow}}\hat{n}_{i,{\downarrow}}
\label{He}
\end{equation}
donde   $\hat c^{\dagger}_{i,{\sigma}}$ ($\hat c^{\phantom{\dagger}}_{i,{\sigma},}$) es el operador de creaci—n (aniquilaci—n) de un electr—n en un sitio $i$ de la banda de conducci—n ($c$) con proyecci—n de esp'n $\sigma\in\{\uparrow,\downarrow\}$. El s'mbolo $\langle i,j\rangle$ se refiere a los sitios primeros vecinos $i,j$. Los operadores de creaci—n y aniquilaci—n obedecen el algebra can—nica de anti-conmutaci—n tal que,
$
[
\hat c^{\phantom{\dagger}}_{i,\sigma},
\hat c^{\dagger}_{j,\nu}
]_{+}=
\delta_{\sigma,\nu}\delta_{i,j}
$,
$
[
\hat c^{\dagger}_{i,\sigma},
\hat c^{\dagger}_{j,\nu}
]_{+}=
0
$
 y 
$
[
\hat c^{\phantom{\dagger}}_{i,\sigma},
\hat c^{\phantom{\dagger}}_{j,\nu}
]_{+}=
0
$,
donde $[A,B]_{\pm}=AB\pm BA$ es el anti-conmutador (conmutador). El operador del nœmero de electrones en cada sitio con proyecci—n $\sigma$  de esp'n esta definido como: $\hat n_{\sigma,i}= \hat c^{\dagger}_{i,{\sigma}}\hat c^{\phantom{\dagger}}_{i,{\sigma}}$. El segundo termino en (\ref{He}) es conocido como el tŽrmino de interacci—n de Hubbard\cite{Hubbard}. El par‡metro $t$ es la amplitud de tunelamiento  y $U$ es la fuerza de interacci—n. El primer tŽrmino de (\ref{He}) tine la interpretaci—n de ser la energ'a cinŽtica efectiva del sistema. En esta interpretaci—n,  los electrones se mueven de sitio en sitio en la red que forma el solido por el potencial efectivo que se genera a partir de los potenciales moleculares efectivos, ver figura \ref{fig1} (a). En una red —ptica se tiene un potencial peri—dico con comportamiento similar dado por las ondas estacionarias generadas por los l‡seres, ver figura \ref{fig1} (b).   Al a–adir la interacci—n electr—nica (tŽrmino con $U$) entender la din‡mica de los electrones es uno de los problemas de mayor interŽs in la materia condensada, ya que se piensa que este modelo contiene los ingredientes esenciales para representar el fen—meno de superconductividad de alta temperatura cr'tica (high-Tc). Al d'a de hoy no existe una teor'a definitiva que explique el mecanismo detr‡s de el fen—meno superconductividad en estos sistemas, llamados cupratos, y otros superconductores ex—ticos. El entender los elementos que controlan la high-Tc es de gran interŽs para el desarrollo de aplicaciones tecnol—gicas y su optimizaci—n\cite{QSim}. 
\begin{figure}[h]
\begin{center}
\includegraphics[width=0.47\textwidth]{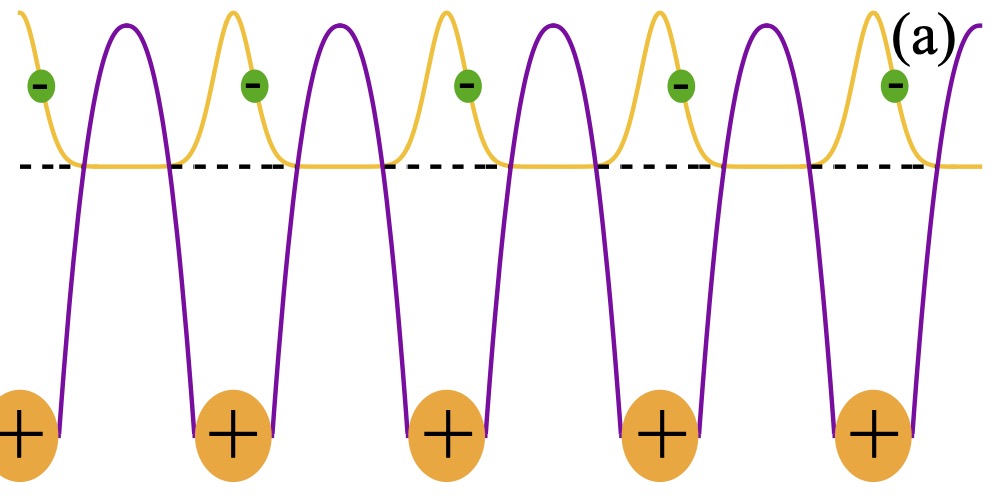}
\includegraphics[width=0.47\textwidth]{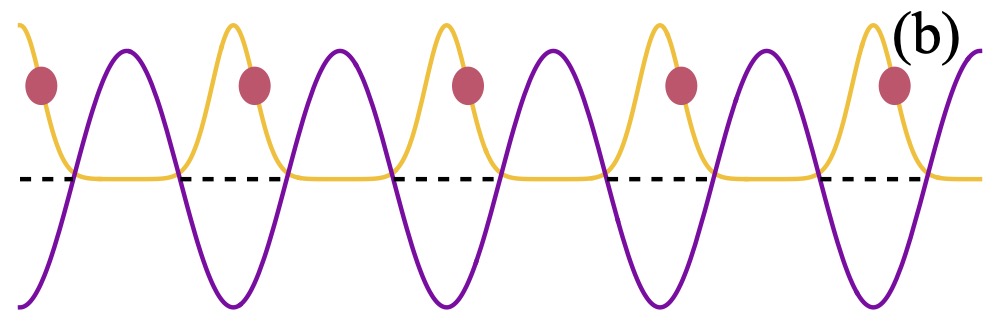}
\end{center}
\caption{(a) Esquema del potencial efectivo del modelo de enlace fuerte de electrones. (b) Esquema ‡tomos ultrafr'os en el potencial efectivo de una red —ptica.}
\label{fig1}
\end{figure}

\subsection{Redes —pticas cl‡sicas}
Las redes —pticas en los sistemas at—micos se puede realizar dado el efecto del campo elŽctrico en los ‡tomos, el efecto Stark \cite{AtomPhys}. Los ‡tomos en un campo elŽctrico tienen una respuesta con un momento dipolar $d_\nu^{(\pm)}$ dado por,
\begin{equation}
d_\nu^{(\pm)}=\sum_{\mu\in\{x,y,z\}}\alpha_{\nu\mu}(\omega_L)E_\mu^{(\pm)}(\mathbf{r},t)
\end{equation} 
donde $\nu\in\{x,y,z\}$, $\omega_L$ es la frecuencia del laser, $\alpha_{\nu,\mu}(\omega_L)$ son los elementos de matriz del tensor de polarizabilidad complejo, $\mathbf{r}$ es la posici—n y $t$ el tiempo. De aqu' se origina que el corrimiento del efecto Stark de corriente alterna (A.C.) cuadr‡tico sirve para desintonizar con respecto a la frecuencia de resonancia at—mica del sistema. El cambio en la energ'a que percibe el ‡tomo es,
\begin{equation}
\Delta E(\mathbf{r},t)=-2\sum_{
\nu,\mu\in\{x,y,z\}}\mathrm{Re}\big[\alpha_{\nu\mu}(\omega_L)E_\nu^{(-)}(\mathbf{r},t)E_\mu^{(+)}(\mathbf{r},t)
\big]
\label{DE}
\end{equation}
de aqu' se deduce que el potencial efectivo, la red —ptica, que ven los ‡tomos es:
\begin{equation}
 V_{\mathrm{OL}}\approx \Delta E\propto \frac{I(\mathrm{r})}{\Delta_a}
\end{equation}
donde $I(\mathrm{r})\propto |E(\mathrm{r})|^2$ es la intensidad del laser y $\Delta_a$ es la desinton'a con respecto a la frecuencia de resonancia at—mica. Esto de manera efectiva es el cambio en la energ'a vista por los ‡tomos debido a los l‡seres asumiendo una respuesta isotr—pica del medio (los ‡tomos). 

\begin{figure}[t!]
\begin{center}
\includegraphics[width=0.34\textwidth]{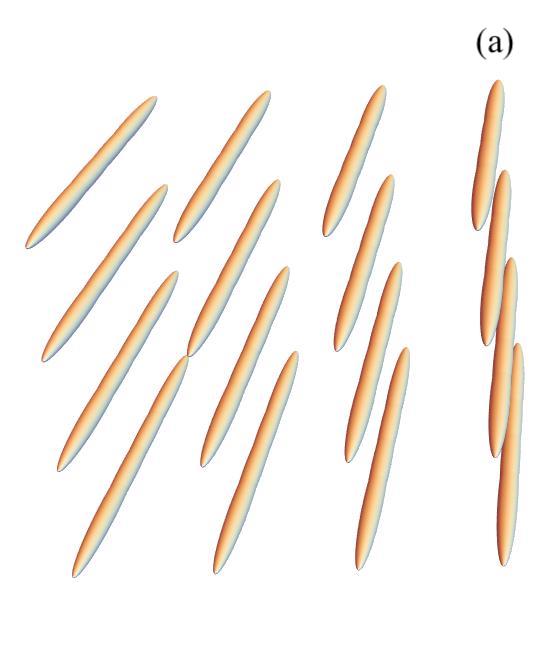}
\includegraphics[width=0.34\textwidth]{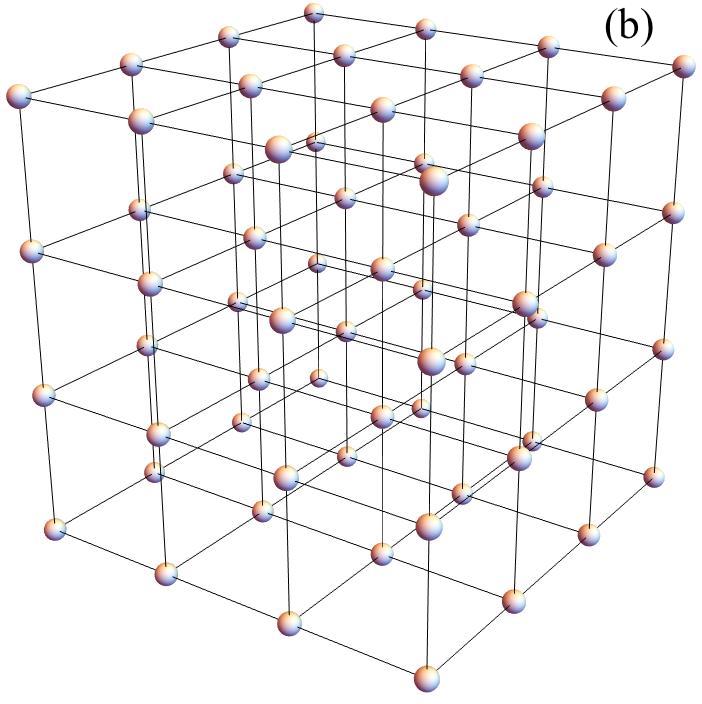}
\end{center}
\caption{(a) Red —ptica de tubos o ``cigarros", los haces contra-propagantes actœan sobre los ejes ``$x$'' y ``$z$'' en 3D. (b) Red —ptica cœbica, los haces contra-propagantes actœan sobre los ejes ``$x$'', ``$y$'' y ``$z$'' en 3D.}
\label{fig1}
\end{figure}
T'picamente uno puede arreglar geomŽtricamente los l‡seres en configuraciones de pares de haces contra-propagantes en diferentes planos espaciales en una, dos, tres dimensiones e incluso otros planos efectivos adicionales arbitrarios\cite{OL, OLSyn1, OLSyn2}. Ejemplos para se muestran en la figura \ref{fig1} (a) para arreglos de tubos o ``cigarros" y (b) una red cœbica. Es posible generar todo tipo de redes peri—dicas, de Bravais e incluso cuasi-cristalinas con suficientes l‡seres. Para una red hiper-cœbica de dimensi—n $d$ uno tiene para el potencial de la red —ptica, 
\begin{equation}
V_{\mathrm{OL}}=\sum_{\nu=1}^dV_0\sin^2(k_\nu x_\nu)
\end{equation} 
donde $k_\nu$ es el vector de onda de la luz en la direcci—n  $\hat e_\nu$, $x_\nu$ es la coordenada (i. e. $x$, $y$, $z$, $\dots$) tal que para para cada dimensi—n $\nu$ tenemos vectores unitarios $\hat{e}_\nu$.  Debido a que el potencial efectivo que obtenemos se obtiene a sin tomar en cuenta aspectos cu‡nticos de la luz en el sistema y los campos de luz est‡n siendo utilizados de manera efectiva paramŽtricamente, nos referiremos a este tipo de redes —pticas como ``redes —pticas cl‡sicas''.

\section{El modelo de Bose-Hubbard}

El modelo de Hubbard de electrones (fermiones), tiene su contraparte en el sistema at—mico en una red —ptica cl‡sica donde se pueden realizar experimentalmente sistemas fermi—nicos ($\phantom{a}^6 $Li, $\phantom{a}^{40}$ K) o sistemas bos—nicos ($\phantom{a}^{87}$Rb). La primera realizaci—n experimental se realiz— con ‡tomos de Rubidio con estad'stica bos—nica. En este caso el modelo de Bose-Hubbard\cite{Lewenstein,Jaksch, Fisher} describe muy bien los experimentos\cite{Greiner}. El modelo se escribe como: 
\begin{equation}
\mathcal{H}^b=-t\sum_{\langle i, j\rangle}(\hat b^\dagger_i\hat b^{\phantom{\dagger}}_j+\mathrm{H.c.})+\frac{U}{2}\sum_i\hat n_i(\hat n_i-1) -\mu\sum_i\hat n_i
\end{equation}
donde   $\hat b^{\dagger}_{i}$ ($\hat b^{\phantom{\dagger}}_{i}$) es el operador de creaci—n (aniquilaci—n) de un ‡tomo en un sitio $i$ con estad'stica bos—nica. Los operadores de creaci—n y aniquilaci—n obedecen el algebra can—nica de conmutaci—n tal que
$
[
\hat b^{\phantom{\dagger}}_{i},
\hat b^{\dagger}_{j}
]_{-}=
\delta_{i,j}
$,
$
[
\hat b^{\dagger}_{i},
\hat b^{\dagger}_{j}
]_{-}=
0
$
 y 
$
[
\hat b^{\phantom{\dagger}}_{i},
\hat b^{\phantom{\dagger}}_{j}
]_{-}=
0
$. 
El operador del nœmero de ‡tomos en cada sitio esta definido como: $\hat n_{i}= \hat b^{\dagger}_{i}\hat b^{\phantom{\dagger}}_{i}$. Los par‡metros del modelo son la amplitud de tunelamiento $t$,  la fuerza de la interacci—n $U$ y el potencial qu'mico $\mu$. 

La modelo tiene dos l'mites simples de analizar. Para $t=0$  (l'mite at—mico) el Hamiltoniano es diagonal.  Para llenado conmensurado el estado base es el de $m N$ ‡tomos distribuidos cada uno en $N$ sitios, donde $m$ es un entero. 
El estado base es simplemente: 
$$
|\psi\rangle_{\mathrm{MI-m}}=|m,m,\dots,m\rangle
$$
En este caso, las fluctuaciones en el nœmero de part'culas, $\Delta n^2=\langle\hat n_i^2\rangle-\langle\hat n_i\rangle^2=0$. Los ‡tomos est‡n completamente localizados. Por ejemplo para una densidad por sitio $\rho=\langle\hat n_i\rangle=m=1$, tenemos un ‡tomo en cada sitio localizado, a este estado se le llama el estado de ``aislante de Mott''  $m$ (Mott-Insulator, MI).

Por otro lado, para $U=0$, el estado base es completamente  des-localizado y presenta coherencia, en este caso se le llama estado ``Superfluido'' (SF). Las fluctuaciones en el nœmero de part'culas por sitio son diferentes de cero y en general pueden ser grandes, del orden del numero de part'culas por sitio, $\Delta n^2\sim\rho$. El estado se puede escribir como,
$$
|\psi\rangle_{\mathrm{SF}}\propto\sum_{q}\sqrt{\frac{M!}{q_1!q_2!\cdots q_N!}}|q_1,q_2,\dots,q_N\rangle
$$
donde $q$ son las posibles combinaciones del conjunto de ocupaciones $q_n$ en $N$ sitios para $M$ ‡tomos.
El cambio de entre estos dos estados al modificar los par‡metros del modelo o en el sistema experimental es un escenario t'pico de una ``transici—n de fase cu‡ntica'' (QPT) \cite{Sachdev}. 
   
\subsection{Aproximaci—n de desacoplamiento, Teor'a de perturbaciones y diagrama de fase}
En esta sub-secci—n seguimos el tratamiento esbozado en \cite{Lewenstein,Stoof} para determinar el diagrama de fase con la aproximaci—n de desacoplamiento y campo promedio.

Para poder tener una estimaci—n de las fronteras de la transici—n de fase cu‡ntica entre el estado superfluido y estado del aislante de Mott (Mott-Insulator)\cite{Fisher}, es necesario realizar aproximaciones. Una aproximaci—n que da buenos resultados cualitativos para este sistema es la aproximaci—n de desacoplamiento en combinaci—n con la aproximaci—n de campo promedio. La aproximaci—n consiste en desacoplar el producto de los operadores de creaci—n y aniquilaci—n en diferentes sitios, suponiendo que las correlaciones cu‡nticas entre ellos son despreciables. Adicionalmente se define un par‡metro de orden, en este caso del orden superfluido. Este par‡metro de orden se construye a partir del valor promedio de los valores de expectaci—n en el estado base de los operadores de creaci—n y aniquilaci—n en el estado base. Esto es,
\begin{eqnarray}
\hat b^{\dagger}_{i}
\hat b^{\phantom{\dagger}}_{j}&\approx&\langle\hat b^{\dagger}_{i}\rangle
\hat b^{\phantom{\dagger}}_{j}+\hat b^{\dagger}_{i}
\langle\hat b^{\phantom{\dagger}}_{j}\rangle-\langle\hat b^{\dagger}_{i}\rangle
\langle\hat b^{\phantom{\dagger}}_{j}\rangle
\\
&=&\psi^*_{i}
\hat b^{\phantom{\dagger}}_{j}+\hat b^{\dagger}_{i}
\psi_{j}-\psi_{i}^*
\psi_{j}
\end{eqnarray}
donde $\psi^*_{i}=\langle\hat b^{\dagger}_{i}\rangle$ y $\psi_i=\langle\hat b^{\phantom{\dagger}}_{i}\rangle$ para un sitio $i$. $\psi$ es el par‡metro de orden superfluido y su significado se aclarar‡ a continuaci—n.

DespuŽs de hacer la aproximaci—n de desacoplamiento en el modelo de Bose Hubbard se puede obtener un Hamiltoniano efectivo por sitio suponiendo que el sistema es homogŽneo y presenta invariancia translacional.
 
\begin{figure}[ht!]
\begin{center}
\includegraphics[width=0.4\textwidth]{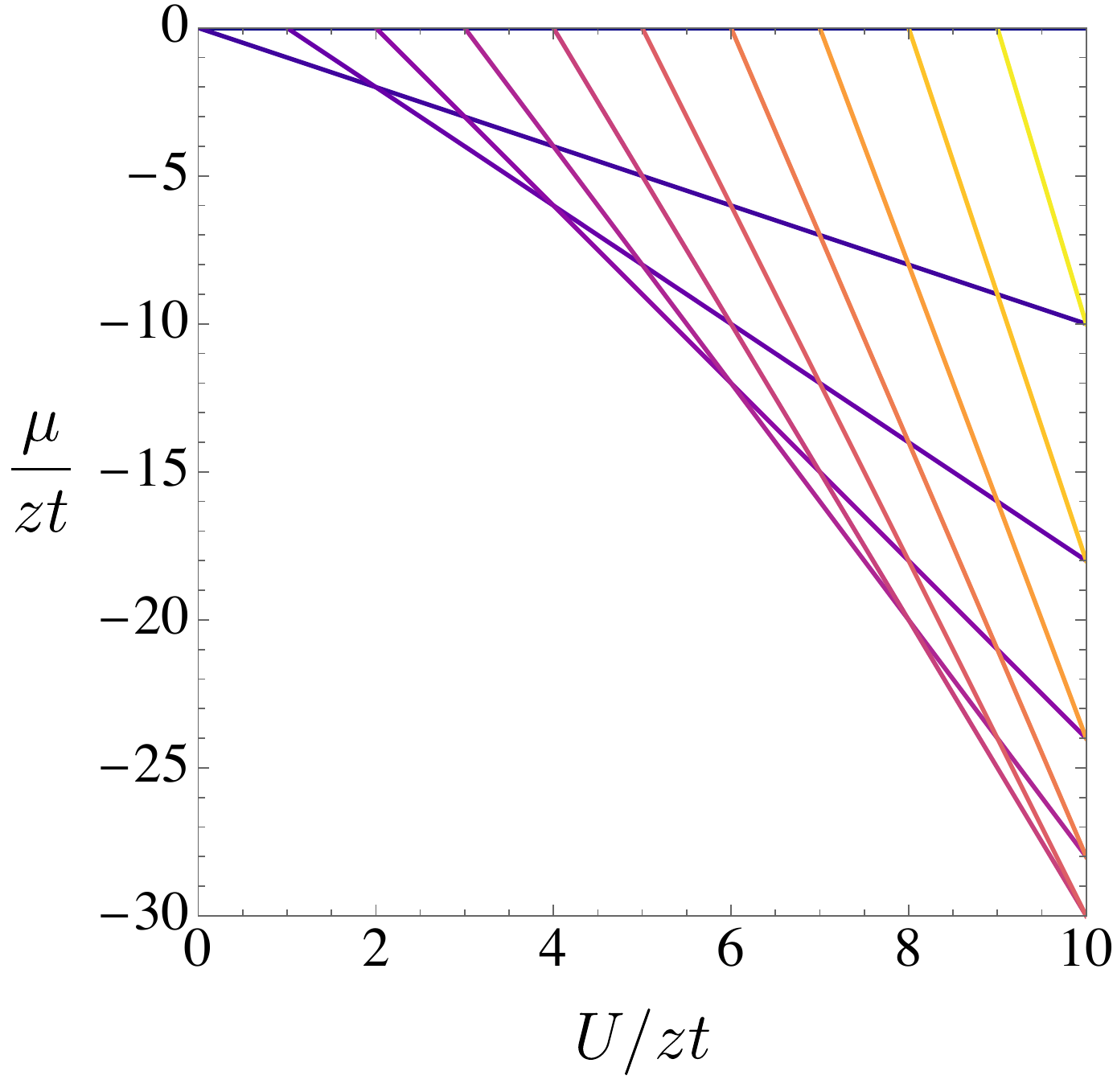}
\end{center}
\caption{Espectro de energ'as de los  estados del Hamiltoniano no perturbado. Las curvas de los estados base como funci—n de $\tilde{\mu}$ se intersectan en los puntos $\tilde{\mu}/\tilde U \in \mathcal{Z} \geq 0$}
\label{F8}
\end{figure}

 El Hamiltoniano por sitio efectivo es:
\begin{equation}
\tilde\mathcal{H}=-\psi(\hat{b}^\dagger+\hat{b})+\psi^2-\tilde{\mu} \hat{n}+\tilde{U}\hat{n}(\hat{n}-1),
\end{equation}
donde $\hat{n}=\hat{b}^\dagger \hat{b}$, $\psi=\langle \hat{b}\rangle$, $\tilde{\mu}=\mu/(zt)$, $\tilde U=U/(zt)$. En lo anterior hemos suprimido el indice por sitio en todas las cantidades  y operadores. Adem‡s, hemos supuesto sin perder la generalidad para este caso que $\psi$ es un nœmero real y hemos supuesto que en promedio es el mismo a lo largo del todo el sistema.
Es conveniente re-escribir el Hamiltoniano de la siguiente manera, 
\begin{equation}
\tilde\mathcal{H}=\tilde{U}\tilde\mathcal{H}_0+\psi \hat{V} +\psi^2,
\end{equation}
donde $\hat{V}=(\hat{b}+\hat{b}^\dagger)$. Es conveniente re-escalar con respecto a $\tilde{U}$, entonces se obtiene:
\begin{equation}
\tilde\mathcal{H}_0=-\frac{\tilde{\mu}}{\tilde{U}}\hat{n}+\hat{n}(\hat{n}-1).
\end{equation}
Adicionalmente, tenemos que $\hat{n}|n\rangle=n|n\rangle$. Debido a que  $\psi^2$ es solo una constante, por el momento no es importante, siendo un corrimiento al espectro de energ'as. Los eigenvalores  (valores propios) de $\tilde{\mathcal{H}}_0$ son:
\begin{eqnarray}
&&\tilde\mathcal{H}_0|g\rangle=E_g|g\rangle\\
&&E_g=0,\quad\frac{\tilde{\mu}}{U}<0\\
&&E_g=-\frac{\tilde{\mu}}{\tilde{U}}g+g(g-1),\quad (g-1)<\frac{\tilde{\mu}}{\tilde U}<g 
\end{eqnarray}
Con la ayuda de la figura \ref{F8} del espectro se puede ver que la relaci—n anterior es v‡lida.
Es posible definir una funci—n simple $\tilde{g}=[\tilde{\mu}/\tilde{U}+1]$, que da como resultado la parte entera de $[\cdot]$. Entonces, se tiene que 
\begin{equation}
E_g=-\tilde{g}\tilde{\mu}/\tilde{U}+\tilde{g}(\tilde{g}-1). 
\end{equation}
Utilizando teor'a de perturbaciones independiente del tiempo no degenerada\cite{Griffiths} y el resultado anterior, se obtiene,
\begin{equation}
E_g(\psi)=\tilde{U} E_g+\psi^2+\psi\underbrace{\langle g|\hat{V}|g\rangle}_{0}+\frac{\psi^2}{\tilde{U}}\sum_{n\neq g}\frac{|\langle g|\hat{V}| n\rangle|^2}{E_g-E_n}+O(\psi^3).
\end{equation}
\begin{figure}[t!]
\begin{center}
\subfloat{\includegraphics[width=0.24\textwidth]{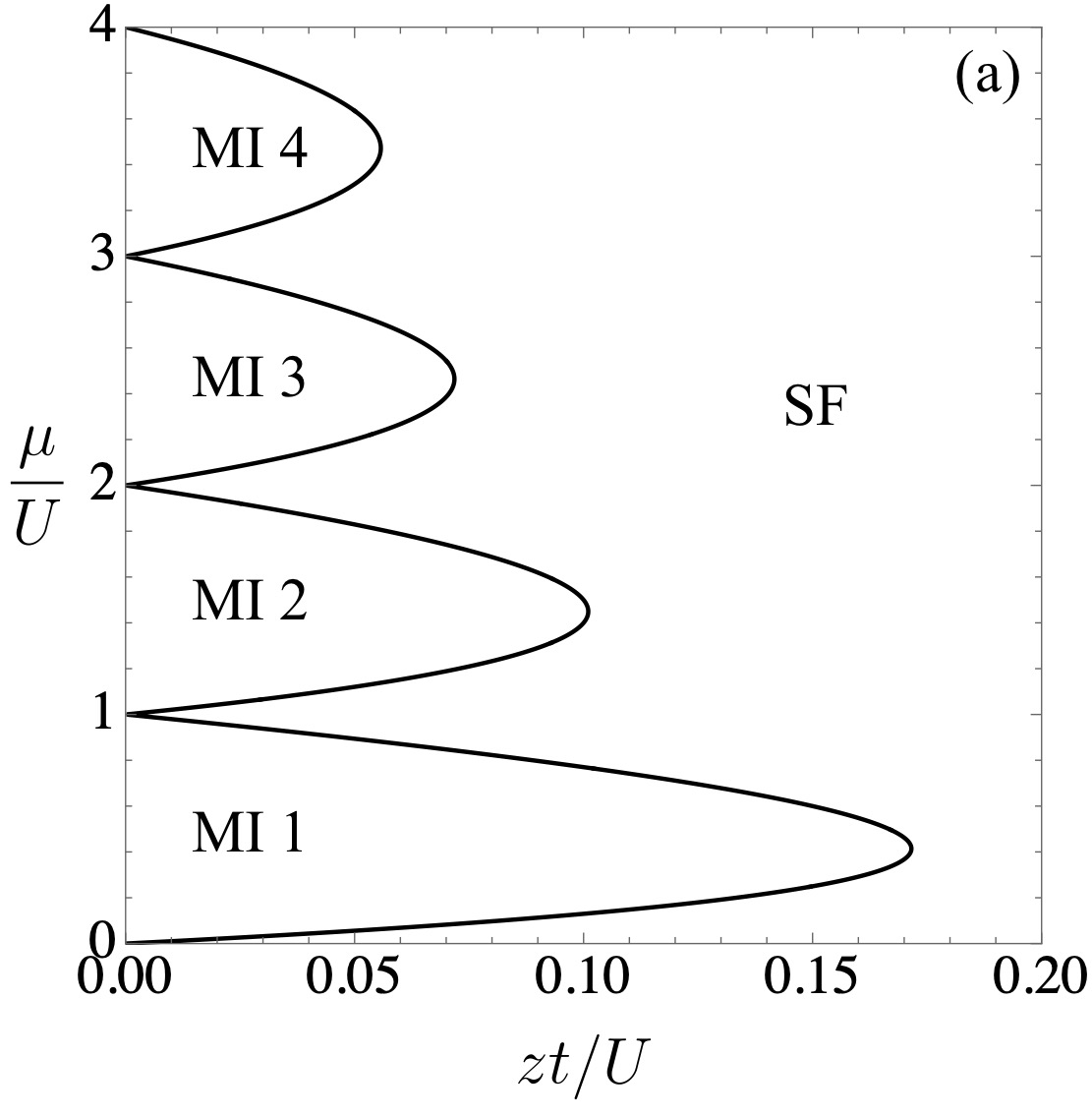}}
\subfloat{\includegraphics[width=0.24\textwidth]{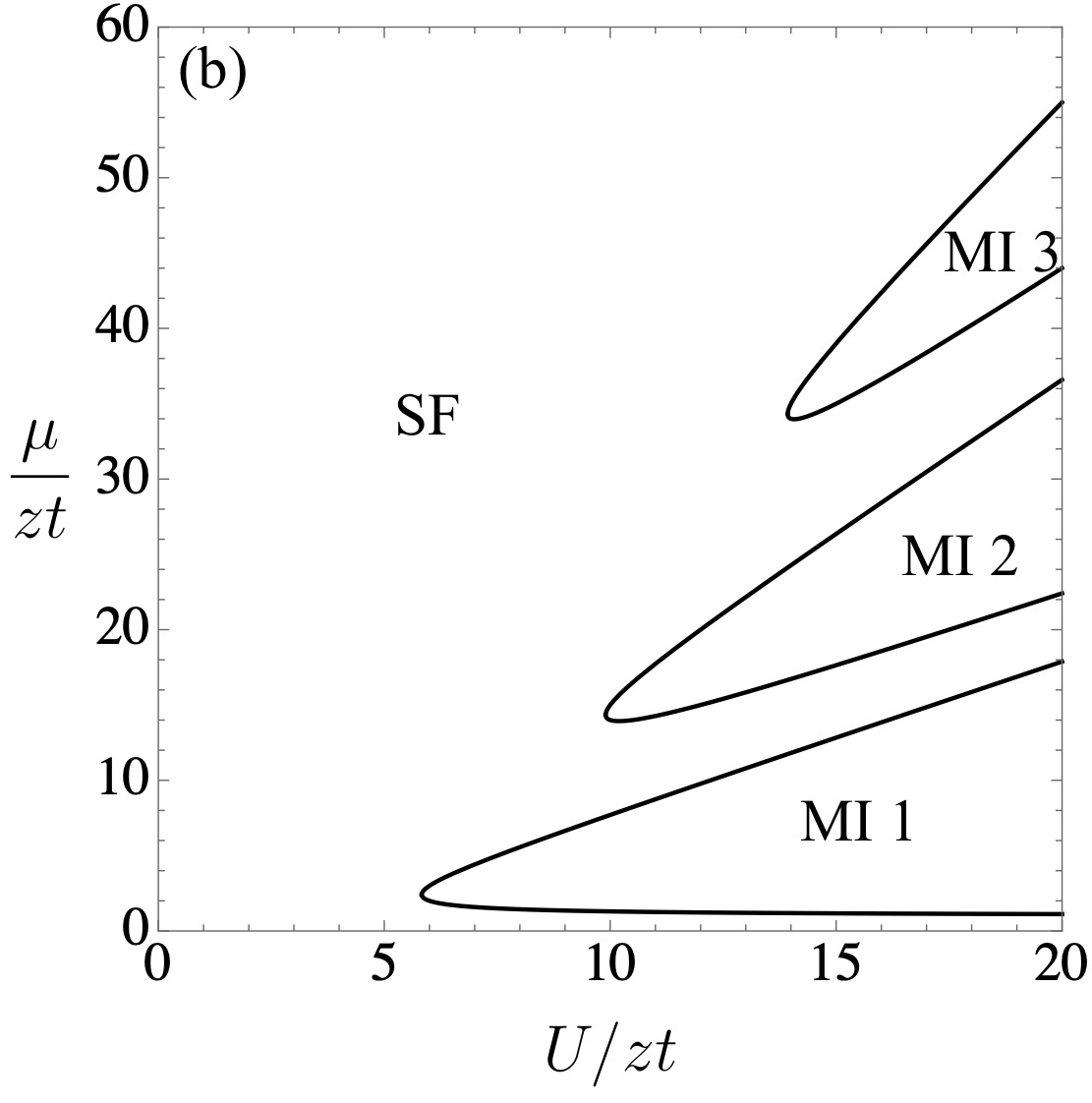}}
\end{center}
\caption{Diagrama de fase del modelo de Bose-Hubbard en aproximaci—n de campo promedio.(a) Presentaci—n t'pica en materia condensada. (b) Presentaci—n t'pica en F'sica at—mica y en experimentos.}
\label{F9}
\end{figure}
Se observa que la correcci—n de primer orden es cero y la correcci—n de segundo orden puede ser f‡cilmente calculada con ayuda de las siguientes definiciones,
\begin{eqnarray}
&&\hat{b} | n \rangle=\sqrt{n}|n-1\rangle\\
&&\hat{b}^\dagger| n \rangle=\sqrt{n+1}|n+1\rangle
\end{eqnarray}
En consecuencia, se obtiene lo siguiente,
\begin{equation}
\sum_{n\neq g}\frac{|\langle g|\hat{V}| n\rangle|^2}{E_g-E_n}=\left(\frac{\tilde{g}+1}{E_g-E_{g+1}}+\frac{\tilde{g}}{E_g-E_{g-1}}\right)
\end{equation}
Utilizando lo anterior se puede escribir lo siguiente,
\begin{equation}
E_g(\psi)=\tilde{U}E_g+\frac{a(\tilde{\mu},\tilde{U})}{\tilde U}\psi^2+O(\psi^4).
\end{equation}
Donde la expresi—n anterior tiene la forma t'pica de un funcional de Landau. Uno puede identificar $\psi$ con un par‡metro de orden y concluir lo siguiente. Si $\psi\neq0$ entonces $a$  debe ser negativo para minimizar la  energ'a. Si de manera contraria, $a$ es positivo, entonces $\psi=0$. Por lo tanto, los valores para los cuales esta transici—n sucede (la linea de transici—n) es controlada por el par‡metro de orden y puede ser encontrada como una funci—n de $\tilde\mu$ y de $\tilde U$. Se tiene que para $\psi\neq 0$ el sistema esta en el estado superfluido y para $\psi=0$ el sistema esta en estado de aislante de Mott. Para que suceda la transici—n se tiene que, 
\begin{equation}
a(\tilde{\mu},\tilde{U})=\tilde{U}+
\frac{\tilde{g}+1}
{\frac{\tilde{\mu}}{\tilde{U}}-\tilde{g}}
+
\frac{\tilde{g}}
{-\frac{\tilde{\mu}}{\tilde{U}}+\tilde{g}-1}=0
,
\end{equation}
Se puede re-escribir la relaci—n de arriba de manera conveniente como:
\begin{equation}
y=\frac{-\tilde{g}^2+2 x \tilde{g} +\tilde{g}-x^2-x}{1+x},
\end{equation}
con $y=1/\tilde{U}$ y $x=\tilde{\mu}/\tilde{U}$. Haciendo una gr‡fica de la funci—n anterior se obtiene el diagrama de fase t'pico del modelo de Bose-Hubbard, ver la  figura \ref{F9}.

\section{Modelos en cavidades de alta reflectancia}

\subsection{Esquemas de cavidades}
\begin{figure}[t!]
\begin{center}
\subfloat{\includegraphics[width=0.24\textwidth]{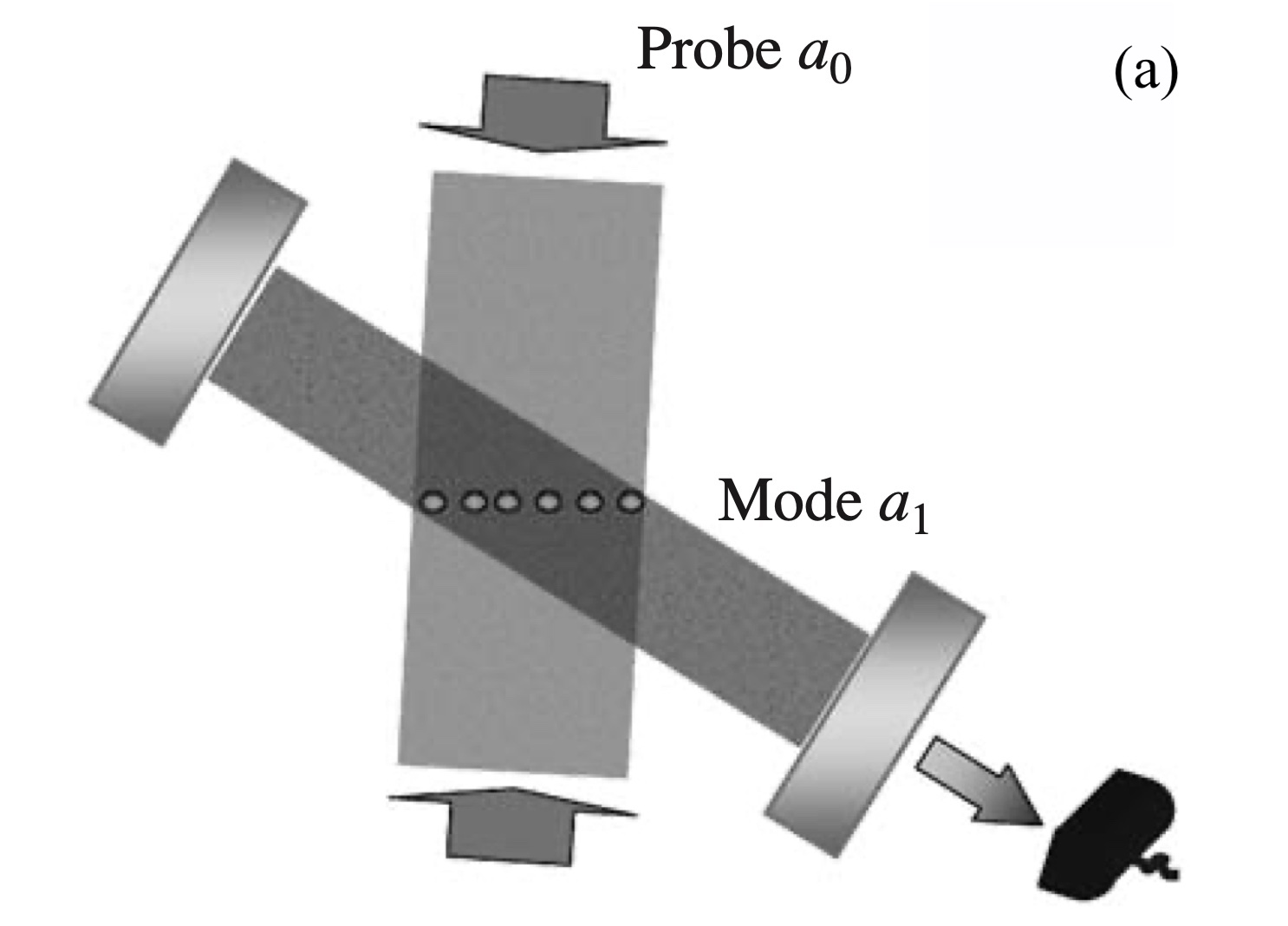}}
\subfloat{\includegraphics[width=0.18\textwidth]{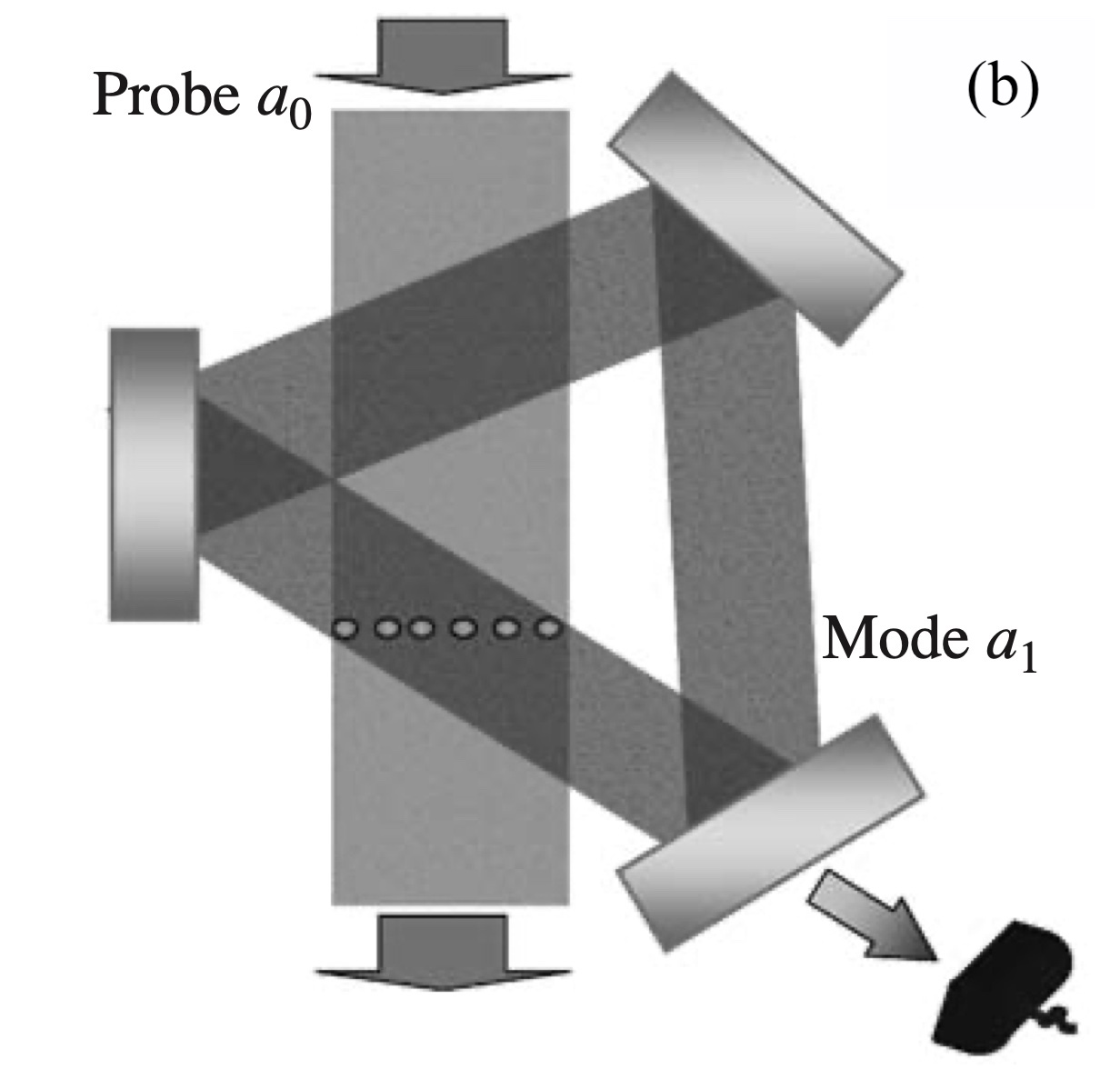}}
\end{center}
\caption{Esquemas experimentales para sistemas de ‡tomos ultrafr'os en cavidades de alta reflectancia (Figuras de \cite{LP9}). (a) Arreglo de cavidad de ondas estacionarias. (b) Arreglo de cavidad de ondas viajeras.}
\label{FC}
\end{figure}

Es posible experimentalmente colocar una cavidad de alta reflectancia en un experimento de ‡tomos ultrafr'os en redes —pticas cl‡sicas\cite{Hemmerich-LS,Esslinger-LS,Esslinger-LS2}. Algunos t'picos arreglos posibles con cavidades de alta reflectancia son los de el esquema de onda estacionaria (a) y onda viajera (b)\cite{LP9}, estos son mostrados en la figura \ref{FC}. Como se puede apreciar en la figura es posible colocar la cavidad en diferentes ‡ngulos con respecto a eje de la red —ptica cl‡sica, es posible colocar diversos detectores de los fotones que escapan la cavidad y se puede hacer incidir luz de bombeo en mœltiples direcciones en principio. Adicionalmente, es posible tener m‡s de una cavidad\cite{Esslinger-CC,Esslinger-CC1} o tener una cavidad multi-modal\cite{Kollar}.

\subsection{Hamiltoniano de luz-materia cu‡ntico}
La materia ultrafr'a dentro de una cavidad de alta reflectancia con una red —ptica cl‡sica se puede representar con el siguiente Hamiltoniano de luz-materia en el rŽgimen cu‡ntico\cite{QOL-Caballero,QSim-Caballero,Bond-Caballero},
\begin{equation}
\mathcal{H}=\mathcal{H}^{l}+\mathcal{H}^{lm}+\mathcal{H}^{m}
\end{equation}
donde $\mathcal{H}^{l}$ es el Hamiltoniano de la luz, $\mathcal{H}^{m}$ es la parte del Hamiltoniano que solo depende de la materia y $\mathcal{H}^{lm}$ es la interacci—n entre la luz y la materia.  A diferencia de las redes —pticas cl‡sicas ahora la naturaleza cu‡ntica de la luz es tomada en cuenta y esta afectar‡ de manera significativa los procesos de correlaci—n que habr‡ en la materia. Expl'citamente los tŽrminos son:
\begin{eqnarray}
\mathcal{H}^{l}&=&\sum_c\hbar\omega_c\hat a_c^\dagger \hat a^{\phantom{\dagger}}_c,
\\
\mathcal{H}^{lm}&=&\sum_{c,p,\sigma} \left(g_{pc\sigma}^*\hat a_c\hat F_{pc\sigma}^\dagger+g_{pc\sigma}\hat a_c^\dagger\hat F_{pc\sigma}\right)
\end{eqnarray}
y $\mathcal{H}^{m}$ es el Hamiltoniano de Bose-Hubbard, Hubbard o incluso puede ser el de una mezcla Bose-Fermi. Los operadores de la luz son $\hat a_c$ donde $c$ es el sub-indice  que denota los modos en la cavidad. Por otro lado el sub-indice $p$ denota los modos de la luz bombeada al sistema y $\sigma$ toma en cuenta la proyecci—n de esp'n correspondiente. Los operadores de luz obedecen relaciones  de conmutaci—n bos—nica. Los acoplamientos de la luz y la materia (i.e. proporcionales a las frecuencias de Rabi) est‡n dados por $g_{pc\sigma}$. Los operadores $\hat F$ denotan la proyecci—n de los modos de luz con respecto a la materia dentro de la cavidad. Los modos que la luz imprimen a la materia pueden ser des compuestos en modos de densidad $\hat D$ y modos de enlace $\hat B$ de la siguiente forma,     
\begin{eqnarray}
\hat F_{pc\sigma}&=&\hat D_{pc\sigma}+\hat B_{pc\sigma}
\\
\hat D_{pc\sigma}&=&\sum_{i}J_{ii}^{pc\sigma}\hat m^\dagger_{i\sigma}\hat m^{\phantom{\dagger}}_{i\sigma}
\\
\hat B_{pc\sigma}&=&\sum_{\langle i,j\rangle}J_{ij}^{pc\sigma}( \hat m^\dagger_{i\sigma}\hat m^{\phantom{\dagger}}_{j\sigma}+\textrm{H.c.})
\end{eqnarray}
donde los operadores $\hat{m}_{i\sigma}^{\dagger}$ pueden corresponder a fermiones $\hat{f}^\dagger$ o bosones $\hat{b}^{\dagger}$.  Adem‡s, las constantes $J_{ii}$ son los traslapes de modos de densidad (tŽrminos en sitio) y las constantes $J_{ij}$ son los traslapes de modos de enlace (tŽrminos a primeros vecinos). Estas constantes de estructura est‡n dadas por las siguientes integrales:
\begin{equation}
J^{pc\sigma}_{ij}=\int w(\mathbf{x}-\mathbf{x}_i)u^*_{c\sigma}(\mathbf{x})u_{p\sigma}(\mathbf{x})w(\mathbf{x}-\mathbf{x}_j)\mathrm{d}^n x
\end{equation}
donde $w(\mathbf{x})$ son las funciones de Wannier de la red —ptica cl‡sica en la que est‡n definidos los operadores de materia que conforman $\mathcal{H}^m$. Tomando en cuenta que las constantes de estructura t'picamente forman un conjunto finito de valores posibles en amplitud con patrones espaciales, es posible definir operadores de modos inducidos por luz tal que,
\begin{eqnarray}
\hat D_\sigma&=&\sum_\varphi J^{D}_{\varphi,\sigma}\hat N_{\varphi,\sigma}
\\
\hat B_\sigma&=&\sum_\varphi J^B_{\varphi,\sigma}\hat S_{\varphi,\sigma}
\end{eqnarray}
con $\varphi$ los modos inducidos por luz (el conjunto de diferentes valores de $J_{ij}$) y los siguientes operadores para estos modos: 
\begin{eqnarray}
\hat N_{\varphi,\sigma}&=&\sum_{i\in\varphi}\hat m^\dagger_{i,\sigma}\hat m^{\phantom{\dagger}}_{i,\sigma}
\\
\hat S_{\varphi,\sigma}&=&\sum_{\langle i,j\rangle\in\varphi}(\hat m^\dagger_{i,\sigma}\hat m^{\phantom{\dagger}}_{j,\sigma}+\mathrm{H.c.})
\end{eqnarray}
T'picamente las funciones de estructura de los modos inducidos por la luz tienen son de la forma (en 1D) $u_{p/c}(x)\propto\cos(2\pi q x/\lambda)$ para cavidades de onda estacionaria, donde $q$ es una constante que depende del arreglo geomŽtrico particular de los haces y espejos en el sistema. Expl'citamente para diferentes configuraciones se pueden encontrar en \cite{Wojciech}. El problema completo con grados de libertad de luz y materia es muy dif'cil de resolver. De ah' que una aproximaci—n habitual es la de eliminar adiab‡ticamente la luz. 

\subsection{Modelos efectivos de materia cu‡ntica en cavidades de alta reflectancia. }

En el marco de rotaci—n de la luz es œtil definir la desinton'a de la frecuencia de la cavidad con la luz de bombeo que se hace incidir en el sistema, $\Delta_c=\omega_c-\omega_p$, donde $\omega_{c/p}$ son las frecuencias de la cavidad y el haz de bombeo. Por simplicidad en lo que sigue consideraremos una cavidad mono-modal y un haz de bombeo. Adem‡s  se considera de manera fenomenol—gica una $\kappa$ que es la tasa de pŽrdida de fotones de la cavidad.  A partir de las ecuaciones de movimiento de los operadores de la luz y la materia, se considera el l'mite del estado estacionario de la luz. Esto es el l'mite donde 
\begin{equation}
\left\langle\frac{d\hat{a}}{dt}\right\rangle \stackrel{!}{=}0\quad\rightarrow\textrm{l'mite adiab‡tico}
\end{equation} 
a partir de esta condici—n, en el l'mite $\kappa\ll|\Delta_c|$,  que significa que la pŽrdida de fotones en el sistema es peque–a o equivalentemente que la cavidad es muy buena, se puede aproximar cualitativamente,
\begin{equation}
\hat a\sim \frac{g\hat F}{\Delta_c+i\kappa}
\end{equation}
Los detalles de diferentes maneras de realizar la aproximaci—n o un tratamiento equivalente se encuentran en \cite{EPJD08,NJPhys2015}. El tratamiento a detalle de la eliminaci—n adiab‡tica lleva al siguiente Hamiltoniano efectivo s—lo con grados de libertad de materia (sin esp'n),
\begin{equation}
\mathcal{H}^m_{\mathrm{eff}}=\mathcal{H}^m+\frac{\hbar\Delta_c|g|^2}{2N_s(\Delta_c^2+\kappa^2)}(\hat F^\dagger\hat F+\hat F\hat F^\dagger)
\label{effmodel}
\end{equation}
donde $N_s$ es el nœmero de sitios en la red —ptica cl‡sica que describe el sistema. Se debe notar que el nœmero de ‡tomos en el sistema es del orden del nœmero de sitios. La luz dentro de la cavidad media una interacci—n en la materia de largo alcance que tiene estructura dictada por como se inyecta la luz al sistema y el arreglo geomŽtrico de la cavidad, as' como la red —ptica. 

{\textbf{Bosones sin esp'n}.} A partir del modelo efectivo en (\ref{effmodel}) es posible considerar el limite donde solamente acoplamiento en las densidades sucede. En la configuraci—n donde el potencial de la red —ptica cl‡sica es lo suficientemente profundo $J_{ij}\approx 0$ para primeros vecinos y bombeando la luz a $90^\circ$ con respecto al eje de la cavidad, $J_{ii}=J_D(-1)^{i}$. Entonces, se puede escribir el siguiente modelo efectivo, 
\begin{equation}
\mathcal{H}=\mathcal{H}^b+\frac{g_{\mathrm{eff}}}{N_s}\left[\sum_\nu (\hat n_{O,\nu}- \hat n_{E,\nu})\right]^2,
\label{QOL}
\end{equation}
donde $\nu$ corre sobre la mitad de los sitios y $g_{\mathrm{eff}}\sim \hbar\Delta_c J_D^2|g|^2/(\Delta_c^2+\kappa^2)$. Este modelo efectivo ha sido realizado experimentalmente\cite{Esslinger-LS,Hemmerich-LS}. Se encuentra que adem‡s de existir las fases de la materia del estado aislante de Mott y estado superfluido, el sistema presenta un estado de aislante de Densidad (DW) con simetr'a  $\mathrm{Z}_2$ rota (Par-Impar) y uno con estado superfluido y de DW coexistiendo, el estado Supersolido (SS). El estado DW puede ser visualizado con el orden que sucede en un tablero de ajedrez donde los ‡tomos se localizan en los sitios pares o en los sitios impares.  C‡lculos de este modelo efectivo son consistentes con estos resultados \cite{QOL-Caballero,QSim-Caballero,Bond-Caballero,Nishant}.

En el caso de que las amplitudes $J_{ij}$ para $\langle i,j\rangle$ dominen el comportamiento por condiciones geomŽtricas\cite{Wojciech}, estados con ``orden de enlace'' (bond order) son posibles. Esto da lugar a d'meros espaciales donde las fases relativas de las ondas de materia por sitio se auto-organizan \cite{QOL-Caballero,QSim-Caballero,Bond-Caballero}.

{\textbf{Fermiones con esp'n}.}
En este caso el Hamiltoniano de materia $\mathcal{H}^m$ se puede escribir como,
\begin{equation}
\mathcal{H}^f=-\sum_{\sigma}\sum_{\langle i , j \rangle }t_{ij}\left(\hat{f}_{i\sigma}^\dagger \hat f_{i\sigma}^{\phantom{\dagger}}+\textrm{H.c.}\right)+U_f\sum_{i}\hat{n}_{i\uparrow}\hat{n}_{i,\downarrow}
\end{equation}
quŽ es el Hamitoniano de Fermi-Hubbard, pero ahora en lugar de electrones tenemos ‡tomos $\hat{f}$ con estad'stica fermi—nica y dos proyecciones de esp'n efectivas. En este caso, se puede considerar que la luz se acopla dependiendo de la polarizaci—n de la luz, derecha o izquierda $R/L$ tal que: $a_{p,\theta}\propto(a_{p,L}e^{i\theta}+a_{p,R}e^{-i\theta})$.  Esto genera el siguiente Hamiltoniano efectivo, eliminando adiab‡ticamente la luz\cite{FQOL-Caballero},
\begin{equation}
{\mathcal{H}}^f_{\mathrm{eff}}={\mathcal{H}}^f+\frac{g_{\mathrm{eff}}}{2 N_s}(\hat{D}^\dagger_\theta \hat{D}_\theta^{\phantom{\dagger}}+\hat{D}_\theta^{\phantom{\dagger}} \hat{D}^\dagger_\theta)
\label{FQOL}
\end{equation}
En (\ref{FQOL}), se ha considerado que la red —ptica es lo suficientemente profunda, se bombea luz a $90^\circ$ y solo tŽrminos de densidad son relevantes, 
\begin{equation}
\hat{D}_\theta=\sum_{j} (-1)^j(\hat{n}_{j,\uparrow}e^{i\theta}+\hat{n}_{j,\downarrow}e^{-i\theta})
\end{equation}
donde $\theta$ es el ‡ngulo de polarizaci—n. Este Hamiltoniano puede soportar adem‡s de estados de pares superfluidos (SF), aislantes anti-ferromagnŽticos (AFI) y estados de pares con simetr'a $\mathrm{Z}_2$ rota formando una DW. El estado con DW adem‡s puede tener fracci—n superfluida de pares no nula, dando lugar a al estado de ``onda de densidad de pares" (pair density wave PDW). Es posible hacer ingenier'a de estados manipulando el ‡ngulo de polarizaci—n, el factor de llenado y explorar diferentes estructuras en la competencia de estados SF, AFI, PDW y DW. Adem‡s es posible manipular el diagrama de fases generando estructuras similares a las de sistemas que presentan high-Tc, en la l'nea cr'tica de $T=0$ \cite{FQOL-Caballero}. 

{\textbf{Bosones con esp'n}. Recientemente, el Hamiltoniano (\ref{QOL}) ha sido extendido y estados con propiedades magnŽticas no triviales han sido explorados \cite{SQOL-Caballero}.

\section{Conclusiones}

La luz en el l'mite cu‡ntico induce interacciones efectivas que cambian significativamente el panorama de las fases de materia cu‡ntica de muchos cuerpos en el sistema. Es posible dise–ar por efectos de retro-acci—n por cavidades estados de materia cu‡ntica emergente con correlaciones fuertes. Usando las propiedades de la luz es posible inducir la competencia de ordenes en sistemas de materia cu‡ntica y utilizarlos para hacer simulaci—n cu‡ntica avanzada. Los potenciales cu‡nticos inducidos por la luz ofrecen una oportunidad œnica para innovar buscando nuevos sistemas de control cu‡ntico e ir m‡s all‡ de an‡logos de sistemas de materia condensada.

\end{document}